\documentclass[iop, apj, twocolappendix, numberedappendix]{emulateapj}
\usepackage{ascmac,amsmath,amsthm,amssymb,aas_macros}
\usepackage{framed} 
\usepackage{graphicx}
\usepackage{epstopdf}
\usepackage{mathrsfs}
\usepackage[title]{appendix}
\usepackage{color}
\usepackage{bm}
\usepackage{ulem}
\usepackage[colorlinks=true,urlcolor=black,linkcolor=blue,citecolor=blue]{hyperref}
\usepackage{fancybox}
\newcommand{\bfx}{\boldsymbol{x}}

\newcommand{\bfk}{\boldsymbol{k}}

\newcommand{\deltag}{\delta_{\rm g}}
\newcommand{\deltam}{\delta_{\rm m}}

\newcommand{\dA}{d_{\rm A}}
\newcommand{\wtc}{\widetilde{C}_1}
\newcommand{\pklin}{P_{\rm lin}}

\begin{document}
\title{Improving geometric and dynamical constraints on cosmology with intrinsic alignments of galaxies}

\author{Atsushi~Taruya\altaffilmark{1,2}${}^*$ and Teppei~Okumura\altaffilmark{3,2}}

\email{ataruya@yukawa.kyoto-u.ac.jp}

\altaffiltext{1} {Center for Gravitational Physics, Yukawa Institute for Theoretical Physics, Kyoto University, Kyoto 606-8502, Japan}
\altaffiltext{2}{Kavli Institute for the Physics and Mathematics of the Universe (WPI), UTIAS, The University of Tokyo, Kashiwa, Chiba 277-8583, Japan}
\altaffiltext{3} {Institute of Astronomy and Astrophysics, Academia Sinica, No. 1, Section 4, Roosevelt Road, Taipei 10617, Taiwan}

\begin{abstract}
We show that the spatial correlation of the intrinsic alignments (IAs) of galaxies, measured in galaxy redshift surveys, offers a precision route to improve the geometrical and dynamical constraints on cosmology. The IA has been treated as a contaminant against cosmological probes such as weak gravitational lensing experiments. However, the large-scale correlation of IAs is expected to follow the coherent large-scale matter inhomogeneities. Here, making use of its anisotropic nature, we show that the large-scale IA correlations help to improve the measurements of the geometric distances and growth of structure. In combination with the conventional galaxy clustering statistics, we find that constraints on equation-of-state parameter for dark energy and Hubble parameter can be tighter than those from the clustering statistics alone by a factor of more than $1.5$. 
\end{abstract}
\thisfancyput(14.8cm,0.5cm){\large{YITP-19-126}}

\section{Introduction}

Mapping the large-scale structure of the universe with galaxy surveys is one of the main science drivers for cosmology. Currently, the key observations are baryon acoustic oscillations \citep[BAO;][]{Peebles_Yu1970,EisensteinHu1998}, and clustering anisotropies due to the redshift-space distortions \citep[RSD;][]{Kaiser1987,Hamilton:1997zq}. Their precision measurements offer an important clue to clarify the nature of cosmic acceleration as well as to probe the gravity on large scales \citep[][for a review]{Weinberg_etal2013}. In doing so, the spatial distribution of galaxies is the major observable, ignoring the individual shapes and orientations. While the orientations of distant galaxy images have been established as a promising tool to measure the weak gravitational lensing \citep{2001PhR...340..291B}, intrinsic alignments (IAs) of galaxies are thought to be a contaminant to be removed in the cosmological data analysis \citep{Heavens_etal2000,Lee_Pen2000,Croft_Metzler2000}. There are numerous works to understand the cosmological impact of IAs, and methods to mitigate the effect have been proposed \citep{Joachimi_etal2015,Troxel_Ishak2015}.

So far, the cosmological application of IAs has attracted less attention, and a limited number of work has been done. Yet, there is growing evidence that the spatial correlation of IAs follows the gravitational tidal fields induced by the large-scale structures, and hence it is expected to contain valuable information. In fact, \citet{Okumura_Jing_Li2009} found that the ellipticity autocorrelation of the SDSS luminous red galaxies (LRG), first detected by \citet{Hirata_etal2007} through the galaxy-ellipticity cross correlation, resembles that of the cold dark matter (CDM) halos in cosmological $N$-body simulations \citep[see also][]{Okumura_Jing2009}. Later, \citet{Blazek_etal2011} has tested the linear alignment (LA) model \citep{Catelan_etal2001,Hirata_Seljak2004}, which relates the IAs to gravitational tidal fields, against the LRG samples, and good agreement was found at large scales \citep[see also][for a detailed comparison with simulations]{Okumura_Taruya_Nishimichi2020}. Furthermore, it has been advocated that the statistics of IAs not only provide a complementary probe \citep{Chisari_Dvorkin2013}, but also offer a clue to the early universe that is even difficult to probe with the galaxy clustering data \citep{Schmidt_Jeong2012,Chisari_etal2014,Schmidt_Chisari_Dvorkin2015,Chisari_etal2016,Kogai_etal2018}. Besides, \citet{Okumura_Taruya_Nishimichi2019} have found the clear BAO features in various statistics related to the IAs \citep[see also][]{Faltenbacher_etal2012}. 

Motivated by these, in this Letter, we clarify the impact of using the IA information, in particular, on cosmological constraints through the measurements of BAO and RSD. We show, for the first time, that combining the IA statistics is beneficial, and significantly tighten the constraints on cosmological parameters, including the equation-of-state (EOS) parameters for the dark energy and the Hubble parameter, by a factor of more than $1.5$, compared to those from the galaxy clustering data alone.

\section{Statistics of IA and galaxy density fields}

The primary focus of this Letter is the spatial distribution of galaxies and their orientations projected onto the sky. While the former is characterized by the fluctuations of number density, denoted by $\deltag(\bfx)$, the latter is quantified by the two-component ellipticity, $(\gamma_+,\gamma_\times)$, defined with the minor-to-major-axis ratio $q$ on the celestial sphere: 
\begin{align}
 \left(
\begin{array}{c}
\gamma_+
 \\
\gamma_\times
\end{array}
\right)(\bfx) \equiv \frac{1-q^2}{1+q^2}
\left(
\begin{array}{c}
\cos(2\phi_x) 
\\
\sin(2\phi_x)
\end{array}
\right).
\end{align}
with $\phi_x$ being the misalignment angle relative to the reference axis. We will below set $q$ to zero for simplicity, which corresponds to the galaxy being assumed to be a line along its major axis \citep{Okumura_Jing_Li2009}. In the weak-lensing measurements, a more convenient characterization of the ellipticity distribution is the rotation-invariant decomposition called E-/B-modes, $\gamma_{\rm E,B}$ \citep{Kamionkowski_etal1998,Crittenden_etal2002}, and these are defined, in Fourier space, by $\gamma_{\rm E}(\bfk)+i\,\gamma_{\rm B}(\bfk) \equiv e^{-i\,2\phi_k}\{\gamma_+(\bfk)+i\,\gamma_\times(\bfk)\}$, where $\gamma_{+,\times}(\bfk)$ are the Fourier counterpart of the ellipticity fields, and $\phi_k$ is the azimuthal angle of the wavevector projected on the celestial sphere, measured from the $x$-axis. Then, we consider the two-point statistics among $\deltag$ and $\gamma_{\rm E,B}$. To quantify the cosmological information encoded in these statistics, we adopt the LA model as mentioned above. In Fourier space, it is given by
\begin{align}
 \left(
\begin{array}{c}
\gamma_+
 \\
\gamma_\times
\end{array}
\right)(\bfk) = -\wtc(z)
\left(
\begin{array}{c}
(k_x^2-k_y^2)/k^2
\\
2k_x k_y/k^2
\end{array}
\right)\,
\deltam(\bfk),
\label{eq:gamma_x+_LA}
\end{align}
with $\wtc$ being the redshift-dependent coefficient \citep{Okumura_Taruya2019}. Here we used the Poisson equation to relate the gravitational potential to the mass density field, $\deltam$. Note that the observable ellipticities are density-weighted, i.e., $(1+\deltag)\,\gamma_{+,\times}$, but at large scales, the term $\deltag\,\gamma_{+,\times}$ is higher order and can be ignored. Then Equation (\ref{eq:gamma_x+_LA}) leads to $\gamma_{\rm B}=0$, and the nonvanishing two-point statistics in Fourier space become the auto-power spectra of the galaxy density and E-mode ellipticity, and their cross power spectrum, which we respectively denote by $P_{\rm gg}$, $P_{\rm EE}$, and $P_{\rm gE}$. In redshift space, where the line-of-sight position of galaxies is determined by the redshift, the observed galaxy density field is affected by the effect of RSD. Furthermore, the ellipticity of galaxies is measured on the celestial sphere normal to the line of sight. Thus, all the power spectra considered here exhibit anisotropies along the line-of-sight direction, and denoting the directional cosine between the wavevector and line-of-sight direction by $\mu$, they are expressed as a function of $k$ and $\mu$. In the linear theory limit, we have \citep[see][for their configuration-space counterparts]{Okumura_Taruya2019}
\begin{align}
P_{\rm gg}(k,\mu;\,z)&=(b_1+f\,\mu^2)^2\,\pklin(k;\,z),
\label{eq:pk_gg}
\\
P_{\rm gE}(k,\mu;\,z)&=- \wtc(z)\,(1-\mu^2)\,(b_1+f\,\mu^2)\,\pklin(k;\,z),
\label{eq:pk_gE}
\\
P_{\rm EE}(k,\mu;\,z)&= \{\wtc(z)\,(1-\mu^2)\}^2 \,\pklin(k;\,z).
\label{eq:pk_EE}
\end{align}
Here, we assume the linear bias relation between the galaxy and matter density fields, and $b_1$ is the coefficient. The quantity $f$ is the linear growth rate defined by $f=d\ln D(a)/d\ln a$ with $a$ and $D$ being, respectively, the scale factor of the universe and linear growth factor, and $\pklin$ is the linear-order matter power spectrum at the redshift $z$.

It should be noted that the BAO is imprinted on $\pklin$, and using its characteristic scale as a standard ruler, the geometric distances to the galaxies at redshift $z$, i.e., the Hubble parameter $H(z)$ and angular-diameter distance $\dA(z)$ are determined via the Alcock-Paczynski effect \citep{Alcock_Paczynski:1979}, which further induces the apparent anisotropies on top of the anisotropic power spectra given above. That is, with the Alcock-Paczynski effect, the projected wavenumbers perpendicular and parallel to the line-of-sight direction, $k_\perp$ and $k_\parallel$, are respectively replaced with $(\dA/d_{\rm A,fid})\,k_\perp$ and $(H/H_{\rm fid})^{-1}\,k_\parallel$, and the power spectra given above are further multiplied by the factor $(H/H_{\rm fid})(\dA/d_{\rm A,fid})^{-2}$ \citep{Seo:2003pu,Taruya:2011tz}, where the quantities with subscript indicate those estimated in a fiducial cosmological model.

\begin{figure*}[tb]

\vspace*{-0.5cm}

\begin{center}
\includegraphics[width=0.55\textwidth]{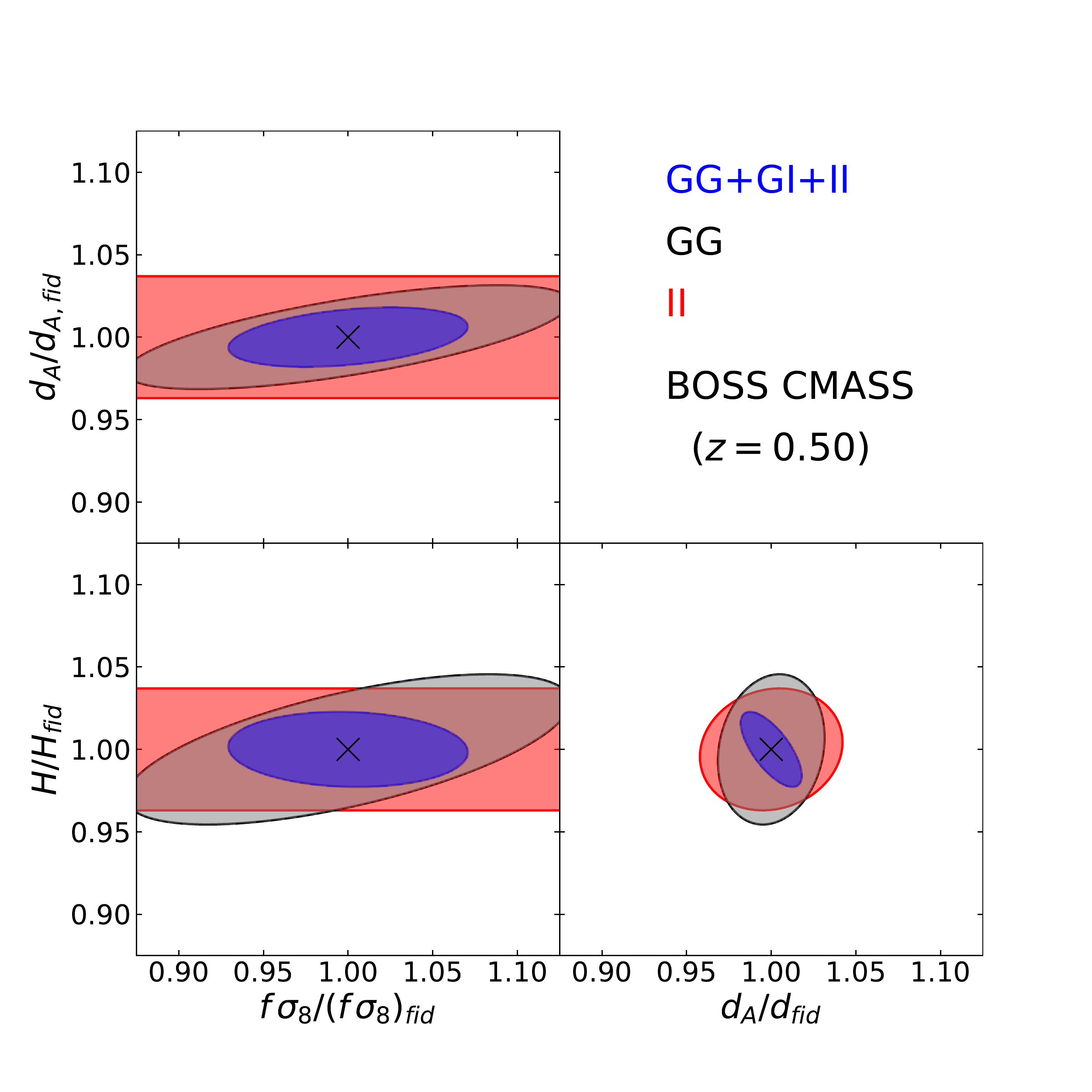}
\hspace*{-1.0cm}
\includegraphics[width=0.47\textwidth]{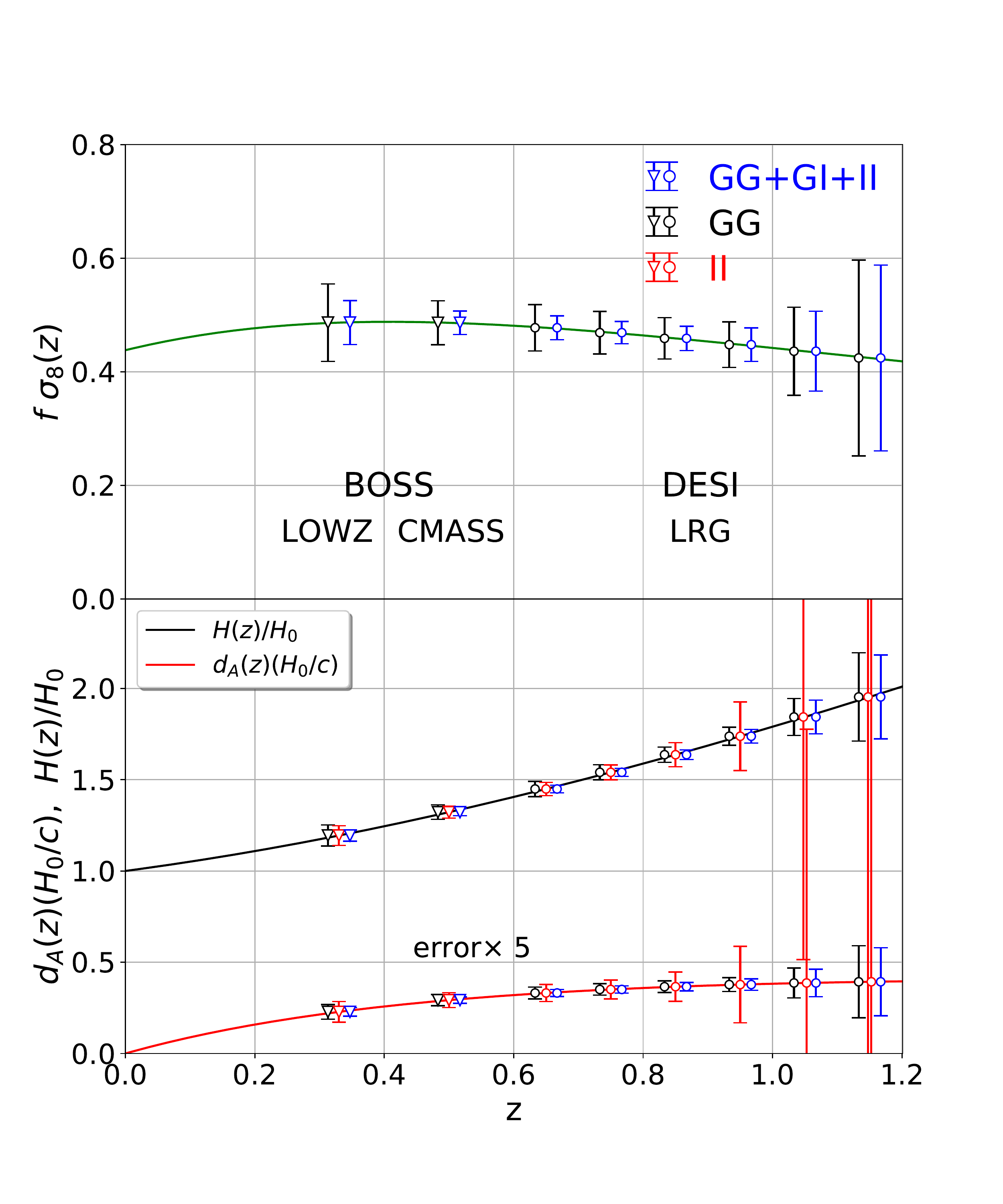}
\end{center}
\vspace*{-0.8cm}
\caption{%
{\it Left :} two-dimensional error contours ($68\%$C.L.) on the geometric distances, $\dA(z)$ and $H(z)$, and the growth of structure, $f\,\sigma_8(z)$, obtained from BOSS CMASS at $z=0.50$. {\it Right :} one-dimensional marginalized errors on the growth of structure (top) and geometric distances (bottom), obtained from BOSS LOWZ ($z=0.33$), CMASS ($z=0.50$) and DESI LRG ($0.6\leq z\leq 1.2$), plotted against the redshift. Solid lines indicate the fiducial model predictions. The errors on $\dA$ are multiplied by $5$ for illustration.
}
\label{fig:Forecasted_constraints_f_da_H}
\end{figure*}

\section{Forecasting cosmological constraints}

Apart from the cosmological information encoded in $\pklin$, the shape and amplitude of the measured spectra $P_a\equiv(P_{\rm gg}$, $P_{\rm gE}$, $P_{\rm EE})$ are characterized by the parameters, $\theta_i\equiv (b_1$, $\wtc$, $f$, $H/H_{\rm fid}$, $\dA/d_{\rm A,fid})$, among which the latter three have explicit cosmological dependencies, and are used to test and constrain cosmological models. To quantify their constraining power, we use the Fisher matrix formalism. Regarding the power spectra $P_a$ as cosmological probes, provided the survey volume $V_{\rm survey}$, minimum and maximum wavenumbers $k_{\rm min}$ and $k_{\rm max}$ for cosmological data analysis, the Fisher matrix is evaluated with 
\begin{align}
& F_{ij}=\frac{V_{\rm survey}}{(2\pi)^2}\int_{k_{\rm min}}^{k_{\rm max}} dk\,k^2\int_{-1}^{1}d\mu 
\nonumber
\\
&\qquad\qquad\qquad
\times\sum_{a,b=1}^3 \frac{\partial P_a(k,\mu)}{\partial \theta_i}
[\mbox{cov}^{-1}]_{ab}
\frac{\partial P_b(k,\mu)}{\partial \theta_j},
\end{align}
where $\theta_i$, the parameters mentioned above, are to be estimated from the measured power spectra. Thus the number of free parameters are five for a given $z$-slice. The matrix $\mbox{cov}_{ab}$ is related to the error covariance of the measured power spectra, whose dominant contributions are the shot noise arising from the discreteness of galaxy distribution, and the cosmic variance due to the limited number of Fourier modes for a finite-volume survey. Focusing on the BAO scales, the Gaussian covariance is a reasonable approximation, and we have
\begin{align}
 \mbox{cov}_{ab} = \left(
\begin{array}{ccc}
 2\,\{\widetilde{P}_{\rm gg}\}^2 &  2\,\widetilde{P}_{\rm gg} P_{\rm gE}& 2\,\{P_{\rm gE}\}^2
\\
 2\,\widetilde{P}_{\rm gg}P_{\rm gE} &  \widetilde{P}_{\rm gg} \widetilde{P}_{\rm EE} + \{P_{\rm gE}\}^2 & 2\,P_{\rm gE} \widetilde{P}_{\rm EE}
\\
 2\,\{P_{\rm gE}\}^2 &  2\, P_{\rm gE} \widetilde{P}_{\rm EE} & 2\,\{ \widetilde{P}_{\rm EE}\}^2
\end{array}
\right),
\label{eq:error_cov}
\end{align}
which is given as a function of $k$ and $\mu$. Here the quantity with tilde is the power spectrum including the shot noise contribution, i.e., $\widetilde{P}_{\rm gg}=P_{\rm gg} + 1/\overline{n}_{\rm gal}$ and $\widetilde{P}_{\rm EE}=P_{\rm EE} + \sigma_\gamma^2/\overline{n}_{\rm gal}$, with $\overline{n}_{\rm gal}$ being the mean number density of galaxies. The quantity $\sigma_\gamma$ represents the scatter in the intrinsic shape per component, including the measurement uncertainty (shape noise). Note that there exist lensing contributions to $P_{\rm EE}$ \citep[e.g.,][]{Matsubara2000lens,Hui_Gaztanaga_Loverde2008}, but we have checked and confirmed them to be ignorable in our setup below.

\section{Setup and results}

Based on the formalism above, we now estimate the constraining power of the IA statistics. For the purpose of illustration, we consider the Baryon Oscillation Spectroscopic Survey (BOSS) LOWZ and CMASS galaxies, which are the largest samples to date at $z\simeq0.33$ and $0.50$. Furthermore, we consider the upcoming survey, Dark Energy Survey Instrument (DESI), and combine its LRG samples at $0.6\leq z\leq 1.2$ with BOSS galaxies to examine how the cosmological parameters are better constrained when combining the IA statistics. Note that with a precision measurement of IAs, we can further extend the analysis up to $z\sim2.4$ \citep{Subaru_PFS2014}. Below, we assume a flat $\Lambda$CDM model determined by \citet{Planck2015_XIII} as our fiducial cosmology. For parameters characterizing the surveys and observed galaxies (i.e., $V_{\rm survey}$, $\overline{n}_{\rm gal}$, and $b_1$), we adopt Table I of \citet{Shiraishi_etal2017} for BOSS samples, and Table 2.3 of \citet{DESI_ScienceBook2016} for DESI LRG samples. To make a conservative estimate, we restrict the analysis to large scales where the linear theory is safely applied, and set $k_{\rm min}$ and $k_{\rm max}$ to $2\pi/V_{\rm survey}^{1/3}$ and $0.1\,h$\,Mpc$^{-1}$, respectively.

The results of the Fisher matrix calculations are shown in Figure~\ref{fig:Forecasted_constraints_f_da_H}, where we separately plot the results using $P_{\rm gg}$ (black), $P_{\rm EE}$ (red), and those using the three power spectra (blue), labeled respectively as GG, II, and GG+GI+II. Here, the redshift-dependent amplitude of E-mode ellipticity $\wtc$ was chosen as $\wtc=c_1/(1+z)^2$ with the fiducial value of $c_1=0.75$, close to the one found in SDSS LRG samples \citep{Okumura_Jing_Li2009,Blazek_etal2011}, setting $q$ to zero. Furthermore, we adopt $\sigma_\gamma=0.3$ for all surveys as a typical shape noise \citep{Schmidt_Chisari_Dvorkin2015}.

The left panel of Figure~\ref{fig:Forecasted_constraints_f_da_H} plots the expected two-dimensional error ($68\%$C.L.) on the growth of structure and geometric distances normalized by their fiducial values, and we specifically show the results from the BOSS CMASS samples. The linear growth rate determined through RSD (i.e., Eqs.~(\ref{eq:pk_gg}) and (\ref{eq:pk_gE})) is known to degenerate with the power spectrum amplitude \citep{Percival:2008sh}, and the actual constraint on the growth rate here is considered in the form of $f\,\sigma_8(z)$, with $\sigma_8$ being the fluctuation amplitude at $8\,h^{-1}$\,Mpc. Clearly, the combination of galaxy clustering data with the IA correlations leads to tighter constraints, and for the CMASS samples, the one-dimensional marginalized error on each parameter is improved by a factor of $1.7-2$, compared to the one obtained from the $P_{\rm gg}$ data alone. This is mainly because the auto-power spectrum $P_{\rm EE}$ is insensitive to the RSD effect. The IA statistics then tighten the constraints on the geometric distances, and this helps break the degeneracy between geometric distances and $f\sigma_8$ through the $P_{\rm gg}$ and $P_{\rm gE}$ data.

These trends are essentially the same for BOSS LOWZ and DESI LRG samples at $z\lesssim0.8$. Right panel of Figure~\ref{fig:Forecasted_constraints_f_da_H} summarizes the one-dimensional marginalized errors on $f\,\sigma_8$ (top), $\dA$ and $H$ (bottom), plotted as a function of $z$. Because of the redshift-dependent amplitude $\wtc\propto(1+z)^{-2}$, the E-mode ellipticity starts to be dominated by the shape noise, and the errors on the geometric distances from $P_{\rm EE}$ data become inflated at $z\gtrsim0.8$. Still, the IA statistics are beneficial, and combining the $P_{\rm EE}$ and $P_{\rm gE}$ data improves the constraint on each parameter by $\sim17\%$ even at $z=0.95$.

\begin{figure}[t]
\begin{center}
\includegraphics[width=0.45\textwidth]{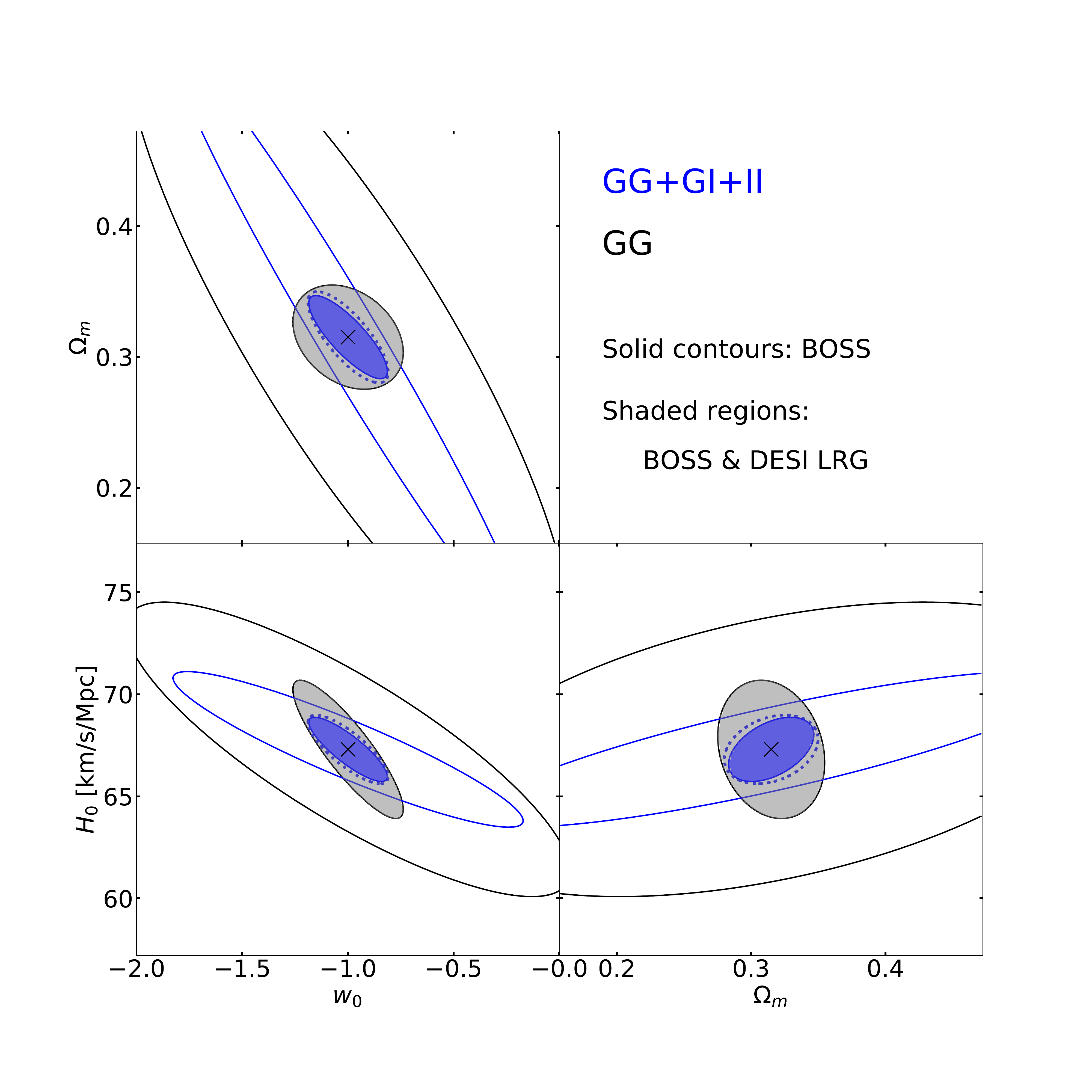}
\end{center}

\caption{%
Joint constraints on cosmological parameters ($w_0, \Omega_{\rm m}, H_0$) from BOSS and DESI, which are obtained by converting the marginalized Fisher matrix for the geometric distances and growth of structure, assuming a flat cosmology. Here, we fix $w_a$, but no prior information is added. In each panel, the error contours ($68\%$ C.L.) on two parameters are plotted, marginalizing over other parameters including $\sigma_8(0)$. The solid lines are the expected errors from BOSS LOWZ and CMASS, while the shaded regions are the combined constraints both from BOSS and DESI LRG. Also, the dotted contours are the combined constraints, but with degraded IA signals from DESI LRG (see the text). The cross symbol in each panel indicates the fiducial value of the cosmological parameters.
}
\label{fig:joint_constraints_BOSS_DESI_4params}

\begin{center}
\includegraphics[width=0.5\textwidth]{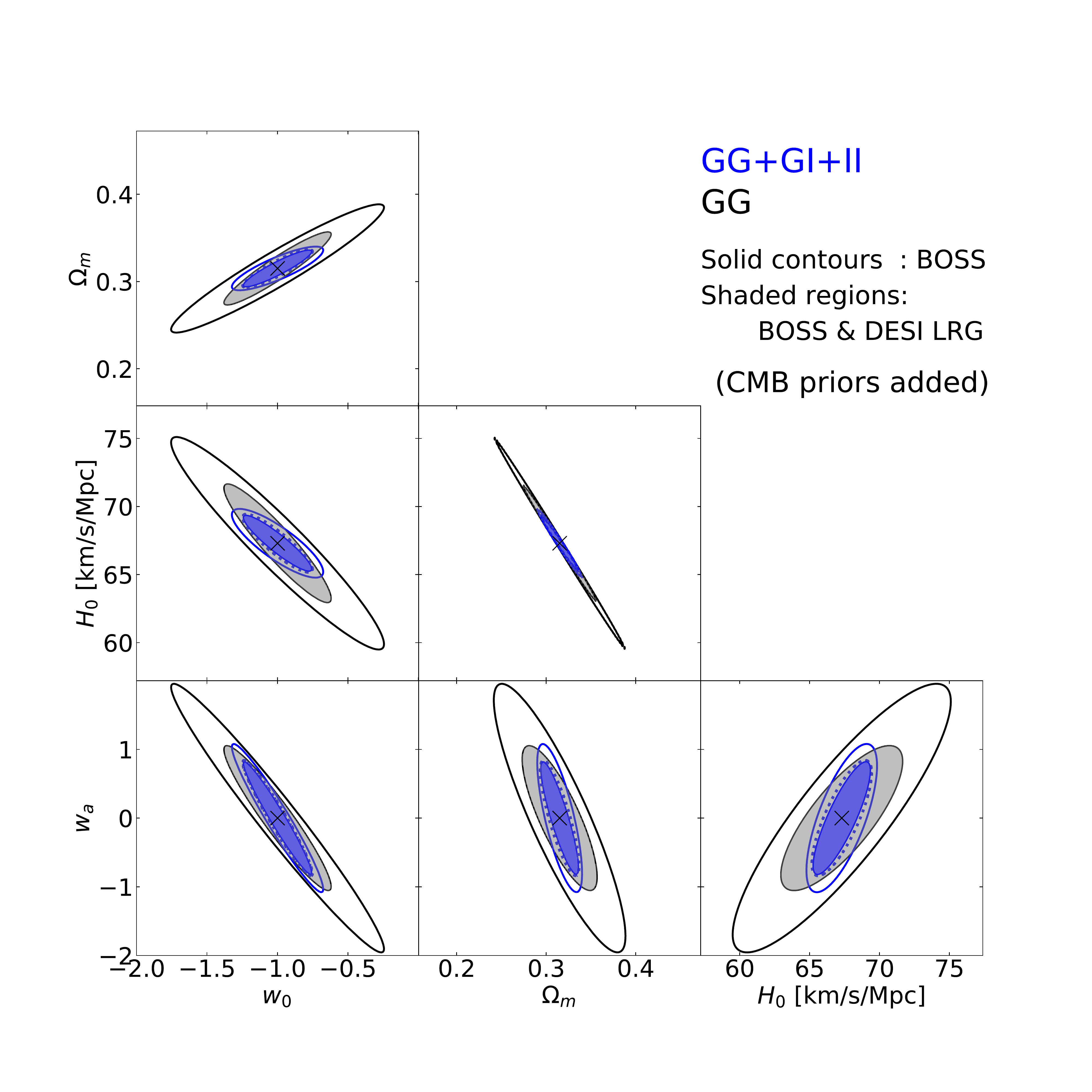}
\end{center}

\caption{%
Same as Figure \ref{fig:joint_constraints_BOSS_DESI_4params}, but the time-varying EOS parameter for dark energy, $w_a$, is allowed to vary. CMB prior information is here added to enhance the scientific impact.
}
\label{fig:joint_constraints_BOSS_DESI_5params}
\end{figure}

Given the model-independent geometric and dynamical constraints in Figure~\ref{fig:Forecasted_constraints_f_da_H}, we can further discuss the specific cosmological model constraints \citep{Seo:2003pu}. As an explicit demonstration, we consider a flat CDM model having the dark energy with the time-varying EOS parameter, $w(a)=w_0+(1-a)w_a$ \citep{Chevallier_Polarski2001,Linder2003}. We then compute the statistical errors on the mass density parameter $\Omega_{\rm m}$, dark energy EOS parameters $w_0$ and $w_a$, and the present Hubble parameter $H_0$, marginalizing over the fluctuation amplitude at the present time, $\sigma_8(0)$. The results are shown as two-dimensional error contours ($68\%$C.L.) in Figures~\ref{fig:joint_constraints_BOSS_DESI_4params} and \ref{fig:joint_constraints_BOSS_DESI_5params}. In deriving the cosmological constraints, surveys at different $z$-slices are assumed to be independent without cross talks.

Figure~\ref{fig:joint_constraints_BOSS_DESI_4params} shows the case for the constant dark energy EOS, fixing $w_a$. Since we do not here use the prior information from the cosmic microwave background (CMB) observations, constraining power on cosmological parameters is restrictive only with the BOSS data. Nevertheless, combining the IA statistics gives a substantial improvement, and the error volume for the three parameters is shrunk by a factor of $5$. Adding the DESI data now gives tighter constraints, and the fractional errors on the Hubble parameter $H_0$ and dark energy EOS parameter $w_0$ are significantly reduced, down to $1.5\%$ and $12\%$, respectively. Although the relative impact of combining the IA statistics is degraded due to the redshift-dependent amplitude $\wtc$, the error volume for the three parameters is reduced by a factor of $3.5$ compared to the one from the galaxy clustering data, thus typically a factor of $1.5$ improvement on each parameter. 

The benefit of combining the IA statistics still holds even when adding the CMB prior information, shown in Figure~\ref{fig:joint_constraints_BOSS_DESI_5params}, where we assume the $0.2$\% and $0.9$\% errors on the determination of CMB acoustic scale and $\Omega_{\rm m}h^2$, respectively. These priors enable us to sufficiently pin down the late-time cosmic expansion, allowing us to constrain the time variation of the dark energy EOS, i.e., $w_a$. Combining the IA statistics, we obtain the one-dimensional marginalized error, $\Delta w_a=0.54$, while the errors on $H_0$ and $w_0$ remain almost the same as those shown in Figure \ref{fig:joint_constraints_BOSS_DESI_4params}. Even with the BOSS data, an excellent performance is expected, and the combination of the IA statistics reduces the error on each parameter by a factor of $1.8-3$. 

\begin{figure}[t]

\begin{center}
\includegraphics[width=0.45\textwidth]{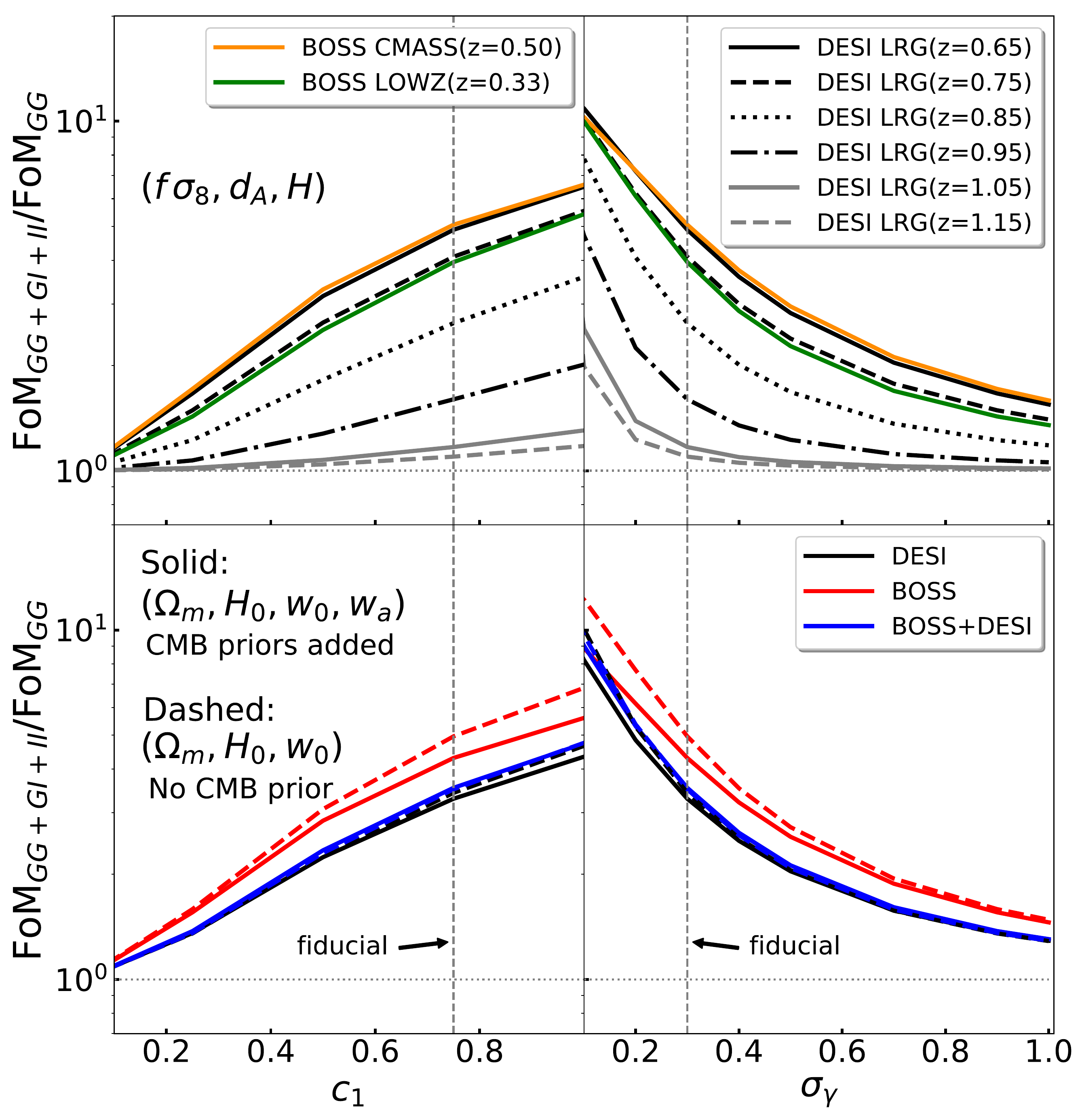}
\end{center}

\vspace*{-0.5cm}

\caption{%
Relative impact of combining the IA statistics on the parameter constraints, defined by the ratio of figure-of-merit, $\mbox{FoM}_{\rm GG+GI+II}/\mbox{FoM}_{\rm GG}$. The results are plotted as functions of the IA parameters, $c_1$ (left) and $\sigma_\gamma$ (right). Upper panels show the results for geometric distances and structure growth, $\dA,\,H,\,f\sigma_8$, derived from each redshift slice of BOSS and DESI. Bottom panels are the results for cosmological parameters, with $\sigma_8(0)$ marginalized over. While the solid lines are the result including CMB priors, the dashed lines are the case without CMB prior information, fixing the time variation of dark energy EOS characterized by $w_a$. }
\label{fig:impact_IA_c1_sigma_gamma}
\end{figure}

\begin{table}[tb]

\vspace*{-0.3cm}

  \caption{Numerical values of the ratio, $\mbox{FoM}_{\rm GG+GI+II}/\mbox{FoM}_{\rm GG}$}
\begin{ruledtabular}
  \begin{tabular}{lccc}
                     &Fiducial& $c_1=0.5$ & $\sigma_\gamma=0.5$ \\ \hline
    BOSS LOWZ        & $3.95$ & $2.53$&  $2.27$\\ 
    BOSS CMASS       & $5.06$ & $3.30$&  $2.95$\\ 
    DESI $(z=0.65)$  & $4.89$ & $3.16$&  $2.83$\\ 
    DESI $(z=0.75)$  & $4.09$ & $2.65$&  $2.39$\\ 
    DESI $(z=0.85)$  & $2.64$ & $1.82$&  $1.68$\\ 
    DESI $(z=0.95)$  & $1.60$ & $1.28$&  $1.23$\\ 
    DESI $(z=1.05)$  & $1.17$ & $1.07$&  $1.06$\\ 
    DESI $(z=1.15)$  & $1.10$ & $1.04$&  $1.03$\\ \hline\hline 
\\ \hline \hline
    Cosmological parameters &Fiducial & $c_1=0.5$ & $\sigma_\gamma=0.5$ \\ \hline
    BOSS                    &  $4.96$ &  $3.07$   &  $2.73$\\ 
    DESI                    &  $3.42$ &  $2.26$   &  $2.05$\\ 
    BOSS+DESI               &  $3.50$ &  $2.30$   &  $2.08$\\ \hline\hline 
\\ \hline \hline

                            &Fiducial & $c_1=0.5$ & $\sigma_\gamma=0.5$ \\ \hline
    BOSS + CMB              &  $4.30$ &  $2.85$   &  $2.56$\\ 
    DESI + CMB              &  $3.29$ &  $2.24$   &  $2.04$\\ 
    BOSS+DESI +CMB          &  $3.53$ &  $2.34$   &  $2.12$\\ 
  \end{tabular}
\tablecomments{Results shown in Figure \ref{fig:impact_IA_c1_sigma_gamma} are tabulated particularly in the cases with $c_1=0.5$ and $\sigma_\gamma=0.5$, together with the results of the fiducial setup ($c_1=0.75$ and $\sigma_\gamma=0.3$, labeled as 'Fiducial'). Upper table shows the results for the BAO and RSD parameters, i.e., $d_{\rm A}(z)$, $H(z)$ and $f\,\sigma_8(z)$, marginalizing over other nuisance parameters. Middle and bottom tables summarize the results for the cosmological parameters with $\sigma_8(0)$ marginalized over, which correspond, respectively, to the dashed and solid lines in the bottom panels of Figure \ref{fig:impact_IA_c1_sigma_gamma}. }
  \label{ratio_FoM}
  \end{ruledtabular}
\end{table}

Note that the outcome of these Fisher matrix analyses relies on our specific setup. In particular, the parameters characterizing the amplitude and error of the measured ellipticity fields, $\wtc$ (or $c_1$) and $\sigma_\gamma$, change the benefit of the IA statistics. To elucidate their impacts, we estimate the figure-of-merit, defined by $\mbox{FoM}\equiv 1/\sqrt{\mbox{det}({\mathcal F}_{ab}^{-1})}$, where ${\mathcal F}_{ab}$ is the sub-matrix of the Fisher matrix for the geometric distances and growth of structure, or that of the converted Fisher matrix for the cosmological parameters, marginalizing over other parameters. Taking the ratio of $\mbox{FoM}$ for the combined data set of galaxy clustering and ellipticity field to that for the galaxy clustering data alone, i.e., $\mbox{FoM}_{\rm GG+GI+II}/\mbox{FoM}_{\rm GG}$, in Figure~\ref{fig:impact_IA_c1_sigma_gamma}, the results for the BAO and RSD parameters (i.e., $\dA$, $H$, and $f\,\sigma_8$) and the cosmological parameters are plotted as functions of $c_1$ (left) and $\sigma_\gamma$ (right). Also, the results with $c_1=0.5$ and $\sigma_\gamma=0.5$ are tabulated in Table \ref{ratio_FoM}, together with those for the fiducial setup. 

As anticipated, the benefit of combining IA correlations largely depends on $c_1$ and $\sigma_\gamma$. For the BAO and RSD parameters, the relative impact varies a lot at low-$z$ slices. Still, we see a sizable improvement on cosmological parameters.  Even with the suppressed amplitude of ellipticity field or enhanced shape noise by a factor of $2$, the relative impact of combining IA correlations exceeds $2$, indicating the $\sim 20\%$ gain for each parameter, compared to the case with galaxy clustering data alone. Figure \ref{fig:impact_IA_c1_sigma_gamma} also indicates that even if the high-$z$ signals of the IA statistics are significantly degraded, combined cosmological constraints are hardly changed. This is explicitly demonstrated in Figures \ref{fig:joint_constraints_BOSS_DESI_4params} and \ref{fig:joint_constraints_BOSS_DESI_5params}, depicted as dotted contours, where smaller values of $c_1$ were chosen for DESI LRG samples, i.e., $c_1=0.5$ at $0.6\leq z\leq 0.8$ and $0.25$ at $0.8<z\leq1.2$.

\section{Conclusion and outlook}

While IAs of galaxies have been considered as the systematics in the cosmological study with weak-lensing observations, their spatial correlation is expected to follow the statistical nature of large-scale structure, and with a proper theoretical modeling, a measurement of galaxy-ellipticity field can deliver the cosmological information, complementary to the galaxy clustering data. We have demonstrated that the large-scale anisotropies in the IA statistics are useful to constrain cosmology, and in combination with the conventional clustering statistics, the IA statistics substantially improve the precision of RSD and BAO measurements, especially at low redshifts. As a result, even restricting the analysis to large scales, the achievable precision from the galaxy surveys at $z=0.3-1.2$ will be improved by a factor of more than $1.5$ for each parameter, including the Hubble parameter and the dark energy EOS parameters. Even reducing the signal of IA correlation by half, the $20\%$ improvement is still possible for the constraint on each cosmological parameter. 

Finally, our forecast results are based on several simplifications and approximations, which have to be verified and/or improved in practical application to observations. Among these, Gaussianity of the error covariance and the linear theory treatment of the RSD ignoring the Fingers-of-God effect \citep{Scoccimarro:2004tg,Taruya:2010mx} are known to respectively change the derived cosmological constraints and the power spectra, although their impacts can be mitigated by restricting the analysis to large scales as we considered here. Another concern would be the accuracy of the LA model to describe the observed ellipticity fields. Albeit its success in good agreement with both observations and simulations, it is the simplest model applicable mainly to elliptical galaxies. Through the observational contamination of other galaxy types as well as possible nonlinear systematics, the use of the LA model may result in a biased parameter estimation. Similar to the galaxy bias \citep{Desjacques_Jeong_Schmidt2018}, the improved theoretical description is indispensable \citep[see, e.g.,][for closely related works]{Blazek_etal2019,Zvonimir_Chisari_Schmidt2019}.

\acknowledgments
This work was supported in part by  MEXT/JSPS KAKENHI grants, Nos. JP15H05889 and JP16H03977 (AT). Numerical computation was partly carried out at the Yukawa Institute Computer Facility. T.O. acknowledges support from the Ministry of Science and Technology of Taiwan under grants No. MOST 106-2119-M-001-031-MY3 and the Career Development Award, Academia Sinica (AS-CDA-108-M02) for the period of 2019 to 2023.


\bibliographystyle{aasjournal}

\end{document}